\documentstyle[preprint,aps,eqsecnum,axodraw]{revtex}

\newcommand{\beq}{\begin{equation}}
\newcommand{\eeq}{\end{equation}}
\newcommand{\beqa}{\begin{eqnarray}}
\newcommand{\eeqa}{\end{eqnarray}}

\def\sm{Standard Model}

\def\Im{\mbox{{\rm Im}}}

\def\avrBR{\langle {\mbox {BR}}\rangle}
\def\BR{\mbox{BR}}

\def\plb#1{Phys.\ Lett.\ {\bf B #1}}
\def\prd#1{Phys.\ Rev.\ {\bf D #1}}
\def\prl#1{Phys.\ Rev.\ Lett. {\bf #1}}

\begin{document}
\draft{\tighten
\preprint{
\vbox{
      \hbox{SLAC-PUB-7713}
      \hbox{hep-ph/9712306}}}

\bigskip
\bigskip

\renewcommand{\thefootnote}{\fnsymbol{footnote}}

\title{Bounding the effect of penguin diagrams in $a_{CP}(B^0 \to \pi^+\pi^-)$}
\footnotetext{Research supported
by the Department of Energy under contract DE-AC03-76SF00515}
\author{Yuval Grossman and Helen R. Quinn}
\address{ \vbox{\vskip 0.truecm}
Stanford Linear Accelerator Center \\
        Stanford University, Stanford, CA 94309 }

\maketitle

\begin{abstract}%

A clean determination of the angle $\alpha$ of the unitary triangle
from $B\to \pi\pi$
decays requires an isospin analysis. If the $B \to \pi^0\pi^0$ and 
$\bar B \to \pi^0\pi^0$ decay rates are small it may be hard to carry 
out this analysis. 
Here we show that an upper bound on the error on $\sin 2\alpha$ due
to penguin diagram effects can be obtained using only the measured
rate $\BR(B^\pm \to \pi^\pm \pi^0)$ and an upper bound on the
combined rate $\BR(B \to \pi^0 \pi^0) + \BR(\bar B \to \pi^0 \pi^0)$. 
Since no $b$ flavor tagging is needed to measure this combined rate, the
bound that can be achieved may be significantly better than any approach
which requires separate flavor-tagged neutral pion information.

\end{abstract}

}
\newpage

\section{Introduction}

The extraction of the angle $\alpha$ of the unitary triangle from
a measurement of the time dependence CP asymmetry in $B \to \pi^+ \pi^-$
is plagued with uncertainty due to penguin diagrams \cite{Yossi}.
This problem can in principle be solved, up to certain discrete ambiguities,
by the Gronau-London isospin analysis \cite{GrLo}, which 
require the measurement of all charge and neutral $B \to \pi\pi$ decays rates.
In practice, however, this theoretically clean determination of
$\alpha$ may be difficult to achieve.
The major problem is expected to be the measurements of
$\BR(B \to \pi^0 \pi^0)$   and  $\BR(\bar B \to \pi^0 \pi^0)$:
The two neutral pion final state is harder to detect and reconstruct
than states with charged pions; furthermore,
arguments base on color suppression \cite{KP} predict a
smaller  branching ratio for this channel than for the two-charged-pion
channel.

Many ways were proposed to disentangle the penguin pollution
in the determination of $\alpha$ \cite{GrLo,allalpha}.
In this note we explain how to set a bound on the error in $\alpha$ 
induced by
the penguin diagram contribution to  the CP asymmetry in 
$B \to \pi^+ \pi^-$.
This bound  requires  the measurement of $\BR(B^\pm \to \pi^\pm \pi^0)$
and only  an upper bound on the combined rate
$\BR(B \to \pi^0 \pi^0) + \BR(\bar B \to \pi^0 \pi^0)$ in addition to the
CP-asymmetry.
The fact that we use only the average of the $B^0$ and $\bar B^0$ rates
removes the need for tagging of these low rate events, 
making this measurement simpler than the measurements of each of the
rate separately.  The error in $\alpha$ decreases when the upper 
bound on the combined rate decreases.
If penguin contributions are large then they may enhance
the $B \to \pi^0 \pi^0$ decay rate. 
If this is so the full isospin analysis can, hopefully, be carried
out. Conversely, if these decay rates are small, the isospin analysis
is difficult, but, as we will show, the uncertainty due to penguin effects in 
the determination of $\alpha$ is small,
since the bound on the penguin diagram contribution is stronger.
In both cases, one can get a meaningful improvement in the knowledge of 
$\alpha$, compared to measuring the charged pion asymmetry alone. 

\section{Derivation of Bounds}

We start with  definitions.
The time-dependent CP asymmetry in $B$ decays
into a final CP even state $f$ is defined as \cite{Yossi}
\beq
a_f(t) \equiv
\frac{\Gamma[B^0(t) \rightarrow f] - \Gamma[\bar B^0(t) \rightarrow f]}
     {\Gamma[B^0(t) \rightarrow f] + \Gamma[\bar B^0(t) \rightarrow f]},
\eeq
and is given by
\beq \label{acp}
a_f(t) = a_f^{\cos} \cos(\Delta Mt) + a_f^{\sin} \sin(\Delta Mt),
\eeq
with
\beq
a_f^{\cos} \equiv {1-|\lambda_f|^2 \over 1 + |\lambda_f|^2}, \qquad
a_f^{\sin} \equiv {- 2 \,\Im \lambda_f \over 1 + |\lambda_f|^2},\qquad
\lambda_f \equiv {q \over p} {\bar A_f \over A_f},
\eeq
where $p$ and $q$ are the components of the interaction
eigenstates in the mass eigenstates,
$A_f(\bar A_f)$ is the $B_d(\bar B_d) \to f$ transition amplitude,
and we will use $|q/p|=1$ \cite{Yossi}.
The time-dependent measurement can separately extract
$a_f^{\cos}$ and $a_f^{\sin}$. In particular,
\beq
\sin[\arg (\lambda_f)] = {\Im\lambda_f \over |\lambda_f|},
\eeq 
can be determined. Notice, however, that
\beq\label{asin}
a_f^{\sin}= {-2\Im\lambda_f \over 1+ |\lambda_f|^2}
= -\sin[\arg (\lambda_f)]\sqrt{1-(a^{\cos}_f)^2}.
\eeq
For $f=\pi^+\pi^-$, and in the absence of penguin diagrams, 
$\lambda=e^{2i\alpha}$ and $a^{\cos}_{+-}=0$ 
(we use $\lambda \equiv \lambda_{+-}$).
Thus, we expect $a^{\cos}_{+-}$ to be small, and difficult to determine
accurately. However
this quantity  only enters quadratically in the correction between
$\sin[\arg(\lambda)]$ and $a^{\sin}_{+-}$.
Thus the error in the value of $\sin2\alpha$ that comes from neglecting this
quantity is small. 
Furthermore, we will show that, for any value of the tree
and penguin contributions, the difference
between $\sin2\alpha$ and $a^{\sin}_{+-}$ is maximized for 
$|\lambda|=1$ ($a_{+-}^{\cos}=0$). 
Hence, our bound is obtained without any dependence
on $a_{+-}^{\cos}$. 

We further define
\beqa
A^{+-} \equiv A(B^0 \to \pi^+ \pi^-), &\qquad& \bar A^{+-} \equiv A(\bar
B^0 \to \pi^+ \pi^-), \\
A^{00} \equiv A(B^0 \to \pi^0 \pi^0), &\qquad& \bar A^{00} \equiv A(\bar
B^0 \to \pi^0 \pi^0), \nonumber \\
A^{+0} \equiv A(B^+ \to \pi^+ \pi^0), &\qquad& \bar A^{-0} \equiv A(B^- \to
\pi^- \pi^0). \nonumber
\eeqa
Isospin symmetry relates these amplitudes
\beq \label{isospin}
{1 \over \sqrt{2}} A^{+-} + A^{00} = A^{+0}, \qquad
{1 \over \sqrt{2}} \bar A^{+-} + \bar A^{00} = \bar A^{-0},\qquad
|A^{+0}| = |\bar A^{-0}|.
\eeq
These equations can be represented by two triangles with unknown
orientation in the complex plane.
Moreover, since the charged final state of two pions $\pi^+\pi^0$ is  a
pure isospin 2 channel
and thus receives no gluon penguin diagram contributions, the CP-conjugate
amplitudes
$A^{+0} $ and $\bar A^{-0}$
are equal in magnitude. (Here we neglect the contribution of electroweak
penguins, since these are at
most at the few percent level \cite{Desh}.) With this approximation
we can draw the two triangles with a common base $|A^{+0}|$, see Fig 1.
This is the Gronau-London construction.

We note in passing that a test for the size of electroweak penguin effects
can be made
by looking for direct CP violation in the $B^\pm \to \pi^\pm \pi^0$ channel
since these  can only
occur due to interference between tree and electroweak penguin terms. While
a null effect could be due to
vanishing relative strong phase between the tree and electroweak penguin
terms, any non-zero effect would be
evidence for enhanced electroweak penguin effects, or possible beyond
standard-model contributions. Hence it is
interesting to search for direct CP violation in this channel, precisely
because it is expected to be small
in the \sm.

Returning to the Gronau-London construction, we further remark that, with
the  common side $A^{+0}$ 
for the two triangles, the angle between  the sides proportional to
$|A^{+-}|$ and $|\bar  A^{+-}|$
is the difference between $\arg(\lambda)$ and $2\alpha$ (see Fig. 1). 
(This is a simple way of stating how the
Gronau-London construction allows extraction of a corrected value of
$2\alpha$.)
Measurement of this angle and of the direct CP violating asymmetry
$a^{\cos}_{+-}$ in addition to the asymmetry
$a^{\sin}_{+-}$ are sufficient to obtain $\alpha$ correctly, 
independent of the
size of the penguin effects.
Likewise, it is now a straightforward matter to 
investigate what constraints on
this angle can be obtained from
a bound on the sum of the  $\pi^0\pi^0$ rates for $B^0$ and $\bar B^0$.

To make this explicit we rewrite Eq. (\ref{asin}) as
\beq\label{deltadef}
a^{\sin}_{+-} = -\sin2(\alpha+\delta)
\sqrt{1-(a^{\cos}_{+-})^2},
\eeq
where we define $2\delta$ as the angle between
$e^{2i\alpha}$ and $\lambda$,
namely  the angle between the $+-$ sides of the two triangles with a common
base $|A^{+0}|$ (see Fig. 1).
A fourfold ambiguity in $2\delta$ arises because
we can flip the orientation of either of the two triangles  about the
common side. For any set of
values of the rates, the larger value of $|2\delta|$, and thus the
largest correction to $\alpha$, occurs
when the two triangles are on the  opposite sides of the base; in our
subsequent derivation of a bound on the
correction  we will consider only this  orientation. Flipping both
triangles about their common side  reverses
the sign of the correction, so  our bound will be on the magnitude of the
correction, with either sign possible.

We define  the combined rate, and the rate ratio
\beq \label{defB}
\avrBR^{00} \equiv {\BR(B^0 \to \pi^0 \pi^0) +  \BR(\bar B^0 \to \pi^0
\pi^0) \over 2}, \qquad
 B^{00} \equiv {\avrBR^{00}\over \BR(B^+ \to \pi^+ \pi^0)}. 
\eeq
We consider what we can learn if $a^{\sin}_{+-}$ and
$\BR(B^+ \to \pi^+ \pi^0)$ are measured, 
and an upper bound on $B^{00}$ is established.
Our goal is to set an upper bound on $|\delta|$.
We emphasis that in what follows we always assume that 
$|\delta|$ is small. Actually, the proofs are correct as long
as $|\delta|<\pi/2$. In practice, of course, we hope to get a much 
tighter bound.

Let us define the angles within the triangles by the labels of the sides
that are opposite them, thus $\phi_{00}$ is the angle opposite to the side of
length $|A^{00}|$, etc. Then, $2\delta = \phi_{00} -\bar \phi_{00}$.
(We use the
pion charges as the labels for the sides, and denote angles in the  
$\bar B$ rate triangle by a bar over the name.)
We  use the sine theorem to write
\beq \label{sines}
\sin \phi_{00} ={|A_{00}| \sin\phi_{+0} \over |A_{+0}|}, \qquad
\sin \bar \phi_{00} ={|\bar A_{00}| \sin\bar \phi_{+0} \over |A_{+0}|}.
\eeq
First we note  that, for a given upper bound on $B^{00}$, 
$|\delta|$ has a maximum.
Geometrically, the  maximum is reached when the two isospin
triangle have opposite
orientation and they are right triangles: 
$|\phi^{+0}|=|\bar\phi^{-0}|=\pi/2$,
so that the sine terms in the right side of Eq. (\ref{sines}) are maximal
(see Fig. 2).
This maximum applies with no knowledge about the values of $|A^{+-}|$ and
$|\bar A^{+-}|$.
Using   Eq. (\ref{sines}) we get
\beq
|\sin\phi^{00}| \le {|A^{00}| \over |A^{+0}|}, \qquad
|\sin\bar\phi^{00}| \le {|\bar A^{00}|\over |A^{+0}|}.
\eeq
Using the fact that
for fixed $x^2 +y^2$, the maximum value of $|x|+|y|$ occurs for $x=y$,
and the definition of $B^{00}$ in Eq. (\ref{defB}) we see that 
$|\delta|$ is maximized when $|\delta|=|\phi_{+0}|=|\bar\phi_{+0}|$,
and obtain
\beq\label{genbound}
 \sin^2\delta \le B^{00}.
\eeq
This is a general bound on $\sin\delta$. 
Note also that in this situation
the two triangles are congruent, which  means that there is no direct CP
violation, when this bound is saturated and thus that the bound on
$\delta$ is achieved when $a^{\cos}_{+-}=0$, as stated above. The known
values of $A^{+-}$ and $\bar A^{+-}$ may constrain the correction to be
slightly less than this generic maximum, but they cannot make it larger.

While we found the maximum value for $\delta$, we still have to 
show that the absolute value of
\beq
\Delta \equiv \sin2\alpha+ a^{\sin}_{+-}=
\sin2\alpha- \sin2(\alpha +\delta)\sqrt{1-|a^{\cos}_{+-}|^2},
\eeq
has a maximum at $|\lambda|=1$.
Note that $|\Delta|$ is symmetric for positive and negative $a^{\sin}_{+-}$.
Moreover, since
$\sqrt{1-|a^{\cos}_{+-}|^2} \le 1$ 
it is clear that for  $|\sin2(\alpha + \delta)|>|\sin2\alpha|$
the effect of $|\lambda| \ne 1$ is to reduce $|\Delta|$.
Thus, we have to check only the case in which 
$|\sin2(\alpha + \delta)|<|\sin2\alpha|$.
In particular, it is enough to check only for
$0 \le \alpha \le \pi/4$ and $-\alpha \le \delta \le 0$.
We differentiate $\Delta$ 
with respect to $|\lambda|$,  taking into account how our
bound on $|\delta|$ is decreased as $|\lambda|$ moves away from 1. 
We find, keeping $|A^{00}|=|\bar A^{00}|$ and
using the geometry of the triangle
construction, that, near $|\lambda|=1$, we can write
\beq
{d\Delta\over d|\lambda |}= (|\lambda|-1)
{\cos(2\alpha + \delta) \over \sin\delta}, \qquad
\left. {d^{\,2}\Delta\over d|\lambda|^2}\right|_{|\lambda|=1} = 
{\cos(2\alpha + \delta) \over \sin\delta}.
\eeq
(In the above we kept $|\lambda| \ne 1$ only when it appears 
in the combination $|\lambda|-1$.)
Thus, we see that the shift in $\sin2\alpha$ is extremal at $|\lambda|=1$.
Since we are concerned with the $\sin\delta<0$ and $\cos(2\alpha + \delta)>0$
case, it is also clear that this is indeed a maximum.

Eq. (\ref{genbound})  is the main result of this note.  
Small improvements to
this result can sometimes be
made if the actual values of $A^{+-}$, $\bar A^{+-}$ and $A^{+0}$ are
inconsistent with
the congruent right triangles  possibility. From the cosine law 
for the two triangles, with a little algebra
and calculus, one can show a more general bound
\beq \label{boundnew}
\sin^2\delta \le 
{(2-\kappa-\bar\kappa)(2B^{00}-\kappa^2-\bar\kappa^2) \over 
4(1-\kappa)(1-\bar\kappa)},
\eeq
where we have defined the ratios
\beq
1-\kappa = {|A_{+-}|\over \sqrt{2} |A_{+0}|} \qquad 
1-\bar \kappa = {|\bar A_{+-}|\over \sqrt{2} |A_{+0}|},
\eeq
and thus $|\lambda|=(1-\bar\kappa)/(1-\kappa)$.
If the neutral rates are too small to measure this is
unlikely to be a
significant improvement over the simpler bound stated above. However, it is
a completely general result, and if $\kappa$ or $\bar \kappa$ are negative
it may provide a slight improvement over Eq. (\ref{genbound}). 

\section{Conclusion}
We wish to stress a few points in our argument 
leading to Eq. (\ref{genbound})
\begin{enumerate}
\item
Since the general bound was saturated for $|\lambda|=1$ it does not require a
measurement of $a^{\cos}_{+-}$.

\item
Since the general bound was obtained for the congruent right triangles case
it does not require measurement
of the actual $B^0$ and $\bar B^0$ to charged pion decay rates, but only
the asymmetry $a^{\sin}_{+-}$, which
reduces sensitivity to errors from cuts that remove backgrounds in the
$B^0$ decay channels.

\item
Finally, since the bound depends only on the sum of the $B^0$ and $\bar
B^0$ decays to neutral pions it can be
determined from untagged data in this channel.
\end{enumerate}
With all these advantages, it is clear that  the bound Eq. (\ref{genbound})
can significantly limit the error on the value of $\alpha$
if a bound on $B^{00} \lesssim 0.1$ can be achieved.

In conclusion, we have shown that a measurements of $\BR(B^+ \to \pi^+\pi^0)$
and an upper bound on the combined $B^0 \to \pi^0\pi^0$ and
$\bar B^0 \to \pi^0\pi^0$ decay rate
can used   to  bound the penguin diagram induced error on the extraction of
$\sin2\alpha$ from the
CP asymmetry, $a_{+-}^{\sin}$,  measurement.  
The bound takes the simple form
\beq
a_{+-}^{\sin} = -\sin2(\alpha+\delta), \qquad
\sin^2\delta \le 
{\BR(B^0 \to \pi^0 \pi^0) +  \BR(\bar B^0 \to \pi^0 \pi^0) \over 
\BR(B^+ \to \pi^+ \pi^0) +  \BR(B^- \to \pi^- \pi^0)}.
\eeq
If the $B$ into neutral pion decay rates are too small to be measured,
then this bound will provide
a determination of the theoretical uncertainty in the value of $\alpha$
extracted from the asymmetry
in two-charged-pion modes without  the assumption of small penguin diagrams
effects.

\acknowledgments
We thank Zoltan Ligeti and Mihir Worah
for helpful discussions. 

{\tighten

}

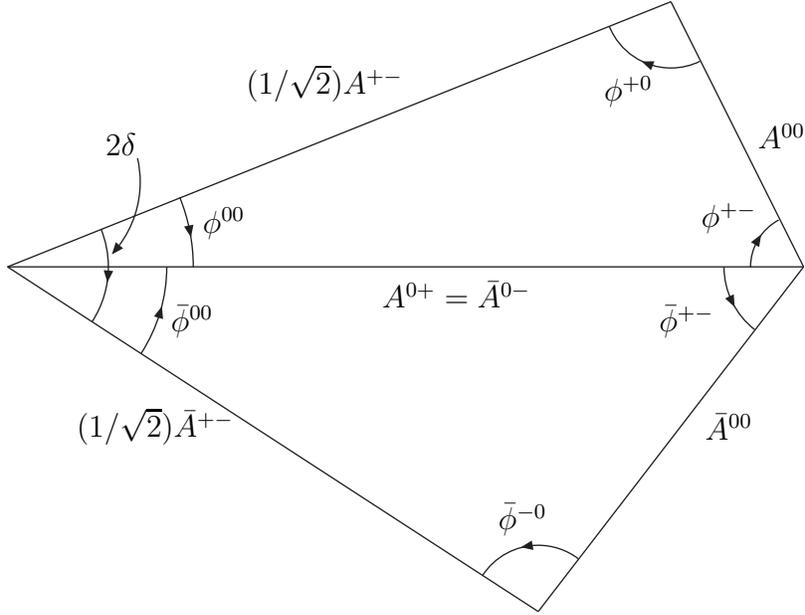
\begin{figure}
\begin{center}
\begin{picture}(400,270)(0,10)
\Line(50,150)(350,150)
\Line(50,150)(250,20)
\Line(50,150)(300,250)
\Line(350,150)(250,20)
\Line(350,150)(300,250)
\Text(220,140)[]{$A^{0+}=\bar A^{0-}$} 
\Text(106,90)[]{$(1/\sqrt{2}) \bar A^{+-}$} 
\Text(170,220)[]{$(1/\sqrt{2}) A^{+-}$} 
\Text(323,90)[]{$\bar A^{00}$}
\Text(343,200)[]{$A^{00}$} 
\ArrowArcn(50,150)(70,22,0)
\ArrowArc(50,150)(60,-33,0)
\ArrowArcn(50,150)(38,22,-33)
\LongArrowArcn(60,182)(40,13,-43)
\Text(93,197)[]{$2\delta$}
\Text(132,167)[]{$\phi^{00}$}
\Text(120,130)[]{$\bar \phi^{00}$}
\ArrowArc(350,150)(30,180,232) 
\ArrowArc(250,20)(25,53,147)
\ArrowArcn(350,150)(20,180,118)
\ArrowArcn(300,250)(25,-64,-158)
\Text(323,169)[]{$\phi^{+-}$}
\Text(307,130)[]{$\bar \phi^{+-}$}
\Text(285,217)[]{$\phi^{+0}$}
\Text(245,55)[]{$\bar \phi^{-0}$}
\end{picture}
\end{center}
\caption[a]{The isospin triangles of Eq. (\ref{isospin}).}
\end{figure}
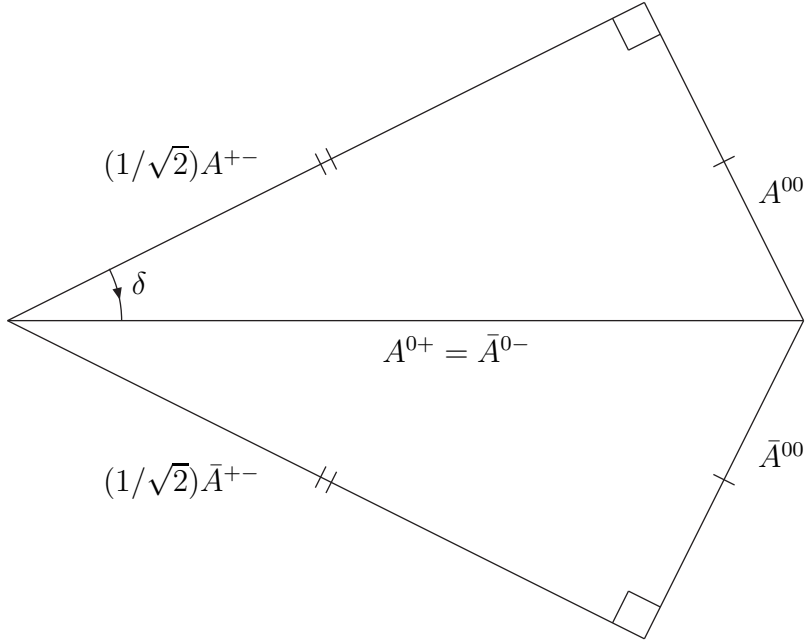
\begin{figure}
\begin{center}
\begin{picture}(400,280)(0,20)
\Line(50,150)(350,150)
\Line(50,150)(290,30)
\Line(50,150)(290,270)
\Line(350,150)(290,30)
\Line(350,150)(290,270)
\Line(278,264)(284,252)
\Line(296,258)(284,252)
\Line(278,36)(284,48)
\Line(296,42)(284,48)
\Line(316,208)(324,212)
\Line(316,92)(324,88)
\Line(166,213)(170,205)
\Line(170,215)(174,207)
\Line(166,87)(170,95)
\Line(170,85)(174,93)
\Text(220,140)[]{$A^{0+}=\bar A^{0-}$} 
\Text(116,90)[]{$(1/\sqrt{2}) \bar A^{+-}$} 
\Text(116,210)[]{$(1/\sqrt{2}) A^{+-}$} 
\Text(343,100)[]{$\bar A^{00}$} 
\Text(343,200)[]{$A^{00}$} 
\ArrowArcn(50,150)(43,27,0)
\Text(100,165)[]{$\delta$}
\end{picture}
\end{center}
\caption[b]{The isospin triangles in the maximum penguin contribution case.}
\end{figure}
\end{document}